\documentclass[sigconf]{acmart}
\usepackage{booktabs} 
\setcopyright{rightsretained}

\copyrightyear{2018}
\acmYear{2018}
\acmConference[IDEAS 2018]{22nd International Database Engineering \& Applications Symposium}{June 18--20, 2018}{Villa San Giovanni, Italy}
\acmBooktitle{IDEAS 2018: 22nd International Database Engineering \& Applications Symposium, June 18--20, 2018, Villa San Giovanni, Italy}
\acmPrice{15.00}

\usepackage{amssymb,amstext,amsmath,amsthm}
\usepackage{tikz}
\usepackage{url}

\DeclareMathAlphabet\mathbfcal{OMS}{cmsy}{b}{n}
\newtheorem{definition}{Definition}[section]

\begin{document}

\title{A Tensor Based Data Model for Polystore}
\subtitle{An Application to Social Networks Data}

\author{\'Eric Leclercq}
\affiliation{%
  \institution{LE2I - EA 7508 - University of Bourgogne\\
  9, Avenue Alain Savary\\
  F-21078}
  \city{Dijon, France}
  \state{}
  \postcode{21078}
}
\email{eric.leclercq@u-bourgogne.fr}

\author{Marinette Savonnet}
\affiliation{%
  \institution{LE2I - EA 7508 - University of Bourgogne\\
  9, Avenue Alain Savary\\
  F-21078}
  \city{Dijon, France}
  \state{}
  \postcode{21078}
}
\email{marinette.savonnet@u-bourgogne.fr}


\begin{abstract}
In this article, we show how the mathematical object tensor can be used to build a multi-paradigm model for the storage of social data in  data warehouses. From an architectural point of view, our approach allows to link different storage systems (polystore) and limits the impact of ETL tools performing model transformations required to feed different analysis algorithms. Therefore, systems can take advantage of multiple data models both in terms of query execution performance and the semantic expressiveness of data representation. The proposed model allows to reach the logical independence between data and programs implementing analysis algorithms. With a concrete case study on message virality on Twitter during the French presidential election of 2017, we highlight some of the contributions of our model.
\end{abstract}




\maketitle

\section{Introduction}
\label{sec:introduction}
 Data from social networks, especially those of Twitter, are increasingly used in applied research projects, in social sciences for example. These data, rich in information about interactions among individuals, allow researchers to understand the digital society's communication models and the interactions between digital social networks, traditional media and citizens.  Results of these researches are relevant to many fields such as marketing, journalism,  public policies study or political communication, as well as reactions to health crises, environmental issues, etc.
However, to address their research questions, social scientists need: 1) to gain control over the data,  namely to contextualize them;  2) to analyze selected data using several algorithms, each puts light on some aspects of the question and; 3) to interpret their results according their knowledge on the subject. For example, the study of political communication on Twitter requires the understanding of viral phenomena, the spread of fake-news and also the role of bots in the dissemination of information.

\begin{figure*}[!h]
\centering
\includegraphics[width=0.7\textwidth]{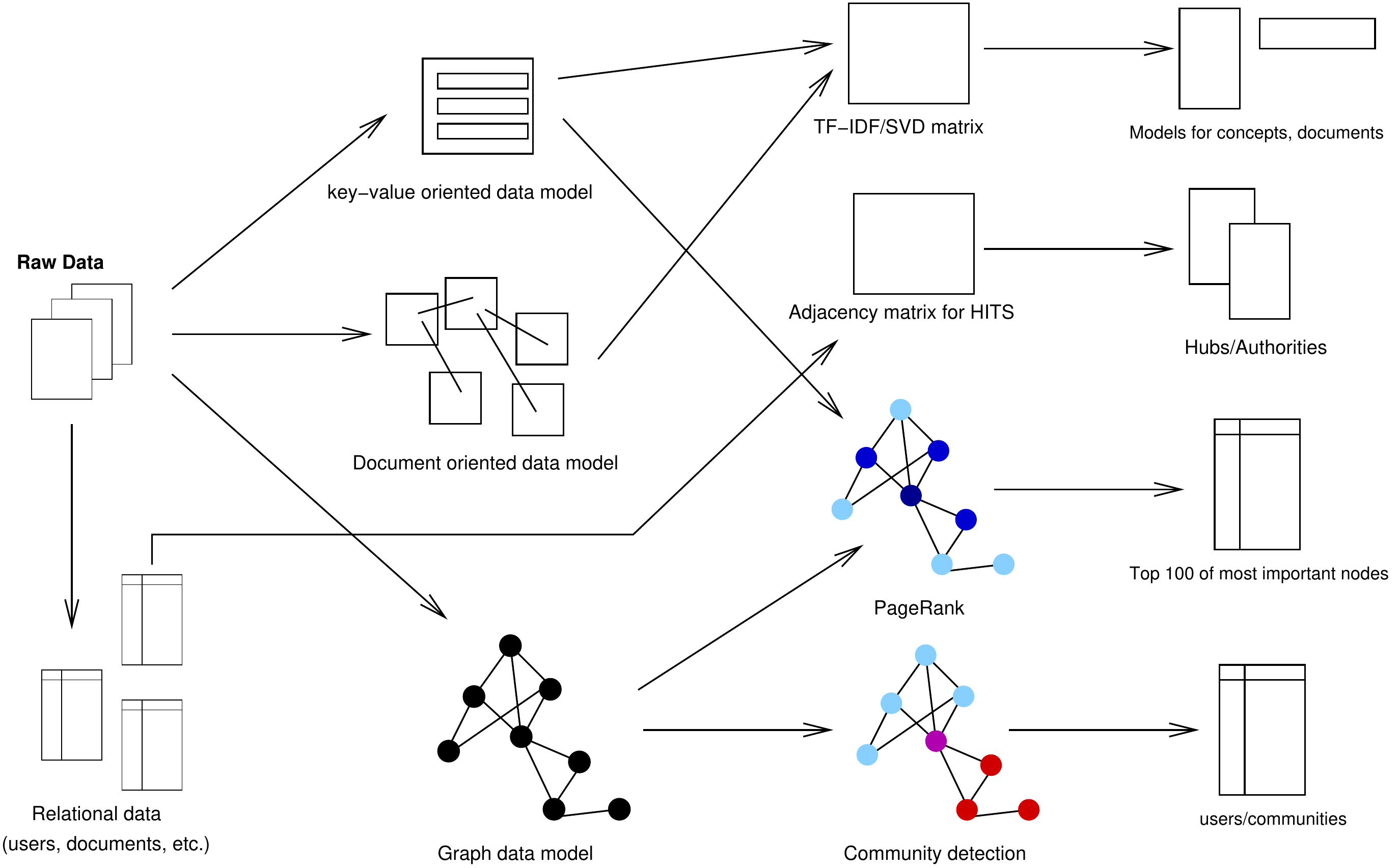}
\caption{Models, models transformations, algorithms.}
\label{fig:principe}
\end{figure*}

Several types of algorithms can be used, for example, to detect communities \cite{Drif2014}, events \cite{Atefeh2015}, influential users \cite{Riquelme2016,AlGaradi2018}, to simulate or study message propagation \cite{Guille2013}. Algorithms hinge on various data models such as graphs, adjacency matrices, multidimensional arrays, time series. In addition,  algorithms do not use the data in the same way, for example graph algorithms can optimize a function and/or perform a random walk on the graph, or detect the edges in a graph through which the number of shortest path between a pair of nodes is the most important (see figure \ref{fig:principe}).

Recent algorithms for social data analysis are rarely implemented in DBMS and matrix operations and associated factorizations (LU, SVD, CUR, etc.) \cite{Leskovec2014,Hogben2013} are not directly supported by storage systems. Only a few NoSQL systems like Neo4j offer quite advanced data graph analysis tools. However, Neo4j does not allow to manage very large amount of data with attributes as the column-oriented systems would do \cite{Holsch2017}. The situation is almost similar for machine learning algorithms and tools. Only some recent systems such as Vertica\footnote{\url{https://www.vertica.com/product/database-machine-learning/}}, SAP HANA platform and its Predictive Analysis Library (PAL)\footnote{https://www.sap.com/community/topic/hana.html},  MLib in Apache Spark\footnote{https://spark.apache.org/mllib/},  or SciDB\footnote{https://www.paradigm4.com/} support standard machine learning algorithms as black-box. Their data models are close to notion of relations and therefore the integration of new machine learning algorithms is complex.  On the other hand some libraries such as Tensor-Flow\footnote{\url{https://www.tensorflow.org/}} or Theano\footnote{\url{http://deeplearning.net/software/theano/}} have been developed to design machine learning tools using data structures close to algorithms. As a results these systems require to develop complex, hard to reuse and often error-prone programs for loading and transforming data.

In recent years we have witnessed the convergence of two separate research fields: High Performance Computing (HPC) and databases towards Data Intensive HPC. One of the concerns of the Data Intensive HPC is to be able to quickly feed the algorithms with  data, as a result, some approaches try to combined several types of storage systems (HDFS distributed file systems, column-oriented databases, etc.) to build an efficient multi-paradigm storage system also called \textit{multistore}, \textit{polystore} or \textit{polyglot storage} \cite{Gadepally2016}. In such systems, data can be partitioned and stored in the model that best fits both the data structure and the algorithms required for their analysis; a partial duplication is also possible. Works on polystores focus mainly on the unification of access by using a common language, few systems propose a model-based approach and  try to achieve physical data independence by using associative array \cite{Kepner2012,Hutchison2017}.

Our objective is to carry out logical data independence in an expressive model that can link models of different data stores while simplifying models transformations and gain performance by leveraging of different systems for computation that there are design to. Our contribution is the definition of a data model based on tensors for which we add the notions of typed schema using associative arrays. We show how the model constructs take place in a mediator/wrapper like architecture. We also define a set of manipulation and analysis operators on tensors.

The remainder of the paper is organized as follows. While section \ref{sec:EA} discusses about OLAP and multi-paradigm storage systems including multistore, polyglot systems and polystores, sections \ref{sec:TDM}, \ref{sec:TDM2} and \ref{sec:op} present the software architecture and the tensor based data model and its operators. Section \ref{sec:exp} presents different experiments and results obtained in context of TEP 2017 project which studies of the use of Twitter by candidates at the french presidential election in 2017.

\section{State of the Art}
\label{sec:EA}
In this section, we describe different approaches to social data warehousing: OLAP (\textit{OnLine Analytical Processing}) data model and multi-paradigm storage.   We end-up this section with a discussion of the strengths and weaknesses of these approaches.

\subsection{OLAP Model and Systems for Social Data}
Since 2010, several works have proposed a multidimensional star schema for building repositories of tweets. Some works are generic and others are directed towards specific analysis.  

In \cite{Bringay2011} the authors have proposed an adapted measure of TF-IDF, where the most significant words  are identified according to levels of dimensions in a cube. Their case study deals with evolution of diseases. Other works \cite{moya2011}, \cite{moalla2014} have built a warehouse dedicated to the sentiment analysis of tweets with a specific schema. 

In \cite{rehman2012,mansmann2014,Kraiem2015}, conceptual models for Twitter data from both OLTP and OLAP point of views are proposed. Models focus mainly on the relationships between tweets and users or among tweets (i.e., retweet, response).

In \cite{Costa2012}, the authors highlight the need to contextualize the data to help analysis. Enrichment requires a formal  knowledge of the domain and is usually dependent on the purpose of analyses. It can be done by linking data to ontology terms and/or by using exploratory analyzes that characterize the data.

In a seminal paper in 2008 \cite{chen2008} the authors describe the \textit{Graph OLAP} approach. They show that traditional OLAP technologies cannot handle efficiently network data analysis because they do not consider links among individual data tuples. They have developed a Graph OLAP framework to define  multi-dimensional and multi-level views over graphs. Given a network dataset with nodes and edges associated to different attributes, a multi-dimensional model can be built so that any portions of a graph can be generalized/specialized dynamically, offering  versatile views of the data. For example, from a citation graph the operation \textit{roll-up} will produce a graph of institutions according to the author dimension.
Favre et al. \cite{favre2017}  propose another approach which consists in enriching the graph by using cubes associated to  nodes and edges.


\subsection{Multi-paradigm Data Storage}
Ghosh states in \cite{Ghosh2010} that storing data the way it is used in an
application simplifies programming and makes it easier to
decentralize data processing. Data storage and processing systems span over several families: relational, NoSQL, array, and distributed file systems. However, transforming various data into a single model may have a significant impact on performance of queries but also on capabilities to apply different algorithms. As stated by Stonebraker in \cite{Stonebraker2005,StoneBraker2007} "one size fits all" is not a solution for modern applications.
As a result, several research projects have been inspired by previous work on distributed databases \cite{Ozsu2011} in order to take advantage of a federation of specialized storage systems with different models\footnote{\url{http://wp.sigmod.org/?p=1629}}. Multi-paradigm data storage relies on multiple data storage technologies, chosen according to the way data is used by applications and/or by algorithms \cite{Sharp2013}. 

In \cite{Tan2017} authors propose a survey of such systems and a taxonomy in for classes:
\begin{itemize}
\item Federated databases systems as collection of homogeneous data stores and a single query interface;
\item Polyglot systems as a collection of homogeneous data stores with multiple query interfaces;
\item Multistore systems as a collection of heterogeneous data stores with a single query interface;
\item Polystore systems as a collection of heterogeneous data stores with multiple query interfaces.  
\end{itemize}

We adopt a slightly different classification based on models and languages by: 1) considering a unique multidatabase query language approach \cite{Litwin1989} instead of federated systems to better represent the autonomy of data sources and pragmatic existing systems; 2) replacing homogeneity of data model systems by isomorphic models\footnote{To be isomorphic two models must allow two way transformations at the structure level but also support equivalence between  sets of operators. For example graph data model and relational data model are not isomorphic because relational data model does not support directly transitive closure.}, for example for JSON and the relational model \cite{Baazizi2017,Discala2016} and; 3) instead of using query interface or query engine terms as a criterion we prefer query language. So our classification is the following: multidatabase query language (unique language), polyglot systems including data models isomorphic to relational model (with multiple languages), multistore, polystore. For each of these classes we describe some of the most significant representatives systems.

Spark SQL\footnote{\url{https://spark.apache.org/sql/}} is the main representative of multidatabase query language. It allows to query structured data from relational like data sources (JDBC, JSON, Parquet, etc.) in Spark programs, using SQL.

According to our classification, CloudMdsQL \cite{Kolev2016} is more a polyglot systems than a multistore system.  CloudMdsQL  is a functional SQL-like language, designed for querying multiple data store engines (relational or NoSQL) within a query that may contain sub-queries to each data store's native query interface. SQL++ which is a part of the FORWARD platform\footnote{\url{http://forward.ucsd.edu/}},  is a  semi-structured query language that encompasses both the SQL and JSON \cite{Ong2015}.

HadoopDB \cite{Abouzeid2009} coupled to Hive\footnote{\url{https://hive.apache.org/}} is a multistore, it uses the map-reduce paradigm to push data access operations on multiple data stores. It can connect to non relational data store such as Neo4j.
D4M (Dynamic  Distributed  Dimensional  Data  Model) \cite{Kepner2012} is a multistore that provides a well founded mathematical interface to tuple stores. D4M allows matrix operations and linear algebra operators composition and applies them to the tuple stores. D4M reduces the autonomy of data stores to achieve a high level of performance \cite{Kepner2014}.

The BigDAWG system \cite{Duggan2015} is a polystore allowing to write multi-database queries with reference to islands of information,
each corresponding to a type of data model (PostgreSQL, SciDB and Accumulo).  
Myria \cite{Wang2017} supports multiple data stores as well as different data computing systems such as Spark. It supports SciDB for array processing, RDBMS, HDFS. The RACO (Relational Algebra COmpiler) acts as a query optimizer and processor for MyriaL language. Myria also supports user data functions in different other languages such as Python.

\subsection{Discussion}
The general limitations of OLAP approaches for social network data are, on one side the performance and on the other side the schema evolution capabilities.

From a performance viewpoint, data models can induce numerous and expensive queries if we need to apply graph algorithms, e.g. to search path with a specific sequence of vertices, compute shortest path between two nodes or for the construction of adjacency matrix. Furthermore, the models transformations are usually expensive, e.g. to transform relational data into a graph or a graph into time series as well as the aggregation of data (\textit{roll-up}) that uses a transitive closure to find all the potential relationships between users linked to an institution using users' citations in tweets. 

From the evolution viewpoint, the issue does not concern data transformation operation but rather the knowledge required to determine the relevant dimensions for analyses. Indeed,  the goal of data modeling for data warehousing is to  build descriptive models to support analyses and then understand  phenomena and produce new knowledge. Usually, social data analysis requires exploratory steps to discover or reveal data properties, to express hypotheses, and then to perform specific analyses on subsets of data, i.e., an iterative approach  producing  incrementally the knowledge. By comparison enterprise data are related to a context including a well defined semantics (sale of product, organization of the company, etc.) while for the social data, the semantic variability is very important. Analyses are not necessarily conducted by the same set of algorithms and each of them can require specific schema. Thus, a data warehouse model must support a set of operators to allow users to define views (materialized or not) close to the shape of the data expected by the algorithms (e.g., adjacency matrix, time series).
For example, although there are few operators (retweets, mentions, hashtags, URLs) in Twitter, and a maximum size of 280 characters per tweet, the datasets generated by Twitter are relatively complex to model and to process: A mention at the beginning of tweet is considered as an inquiry or an answer while multiple mentions at the end of a tweet are interpreted like the desire to expose the tweet to other users (media for example). Retweet is much more complex to interpret because it can involve either a membership or an opposition (for a satiric tweet for example), consequently its interpretation depends clearly on the content and on the context.

Multi-paradigm data storage is less considered in the approaches of data warehousing which stay still mainly in a traditional vision of the DBMS. The  column-oriented or value-oriented NoSQL systems brought another vision which dissociate the features of  DBMS by considering  systems as storage engines for which the usual properties of the RDBMS are elastic. For example, it belongs to the programmer, depending of the NoSQL storage engine, to implement constraints checking in the application layer. The described polystore approaches roughly share the same principle by using a common language to access to storage engine. Some works on scientific data propose a different approach similar to physical data independence by using a generic model based on associative array to subsume relational, graph and matrix models \cite{Hutchison2017,Gadepally2015}. Instead, our approach tackles the logical data independence issue and explores the expressiveness of a model based on the mathematical object tensor.


\section{Overview of Approach and Architecture}
\label{sec:TDM}
Our approach is designed under the following assumption: preserve the local autonomy of the storage systems without considering updates of the data except those consisting in materializing results of models transformations or analyses. It is  a polystore approach  where it is possible to use either the native mode of each system or a tensor-based pivot data model to express queries. Tensor pivot model insures the decoupling between programs and data (\textit{logical data independence}). 

In terms of analysis tools, we have selected two types of languages: R and Spark. Tensor flow  is similar to libraries supporting tensors in R or Spark, but it has been design with a workflow orientation and not with a data model orientation. Thus, tensor is rather a structure of exchange among processes of a complex workflow than a model to represent real data.

In the following sections we will clarify the notion of tensor and describe our polystore architecture but, at first, we use the analogy of a tensor with a multidimensional array or an hypermatrix, that is a family of elements indexed by $N$ sets. 
 
\subsection{From Tensor Mathematical Object to a Data Model}

Tensors are very general abstract mathematical objects which can be considered according to various points of view. Tensors can be seen as multi-linear applications or as the result of  the tensor product.   A tensor is an element of the set of the functions from the product of $N$ sets $I_j, j=1,\dots,N$ to $\mathbb{R}: \mathbfcal{X} \in \mathbb{R}^{I_1\times I_2\times \dots \times I_N}$, $N$ is the number of dimension of the tensor or its order or its mode. 

A tensor is often defined of as a generalized matrix,  0-order tensor is  a scalar, 1-order one is a vector, 2-order one is  a matrix, tensors of order 3 or higher are called higher-order tensors. More formally, a $N$-order tensor is an element of the tensor product  of $N$ vector spaces, each of which has its own  coordinate system. 

Tensor operations, by analogy with the operations on matrices and vectors, are multiplications, transpose, unfolding or matricization and factorizations (also named decompositions) \cite{Kolda2009,Cichocki2009}. The most used tensor products are the Kronecker product denoted by $\otimes$, Khatri-Rao product denoted by  $\odot$, Hadamard product denoted by $\circledast$, external product denoted by $\circ$ and n-mode denoted by  $\times_n$.

In the rest of the article, we use the boldface Euler script letters to indicate a tensor $\mathbfcal{X}$, for matrices the boldface capital letters $\bf M$, the boldface lowercase letters to indicate a vector $\mathbf{v}$, and an element of the tensor or a scalar is noted in italic, for example $x_{ijk}$ is $ijk$-i-th~element of 3-order tensor  $\mathbfcal{X}$. 

\begin{figure*}[!h]
\centering
\includegraphics[width=0.7\textwidth]{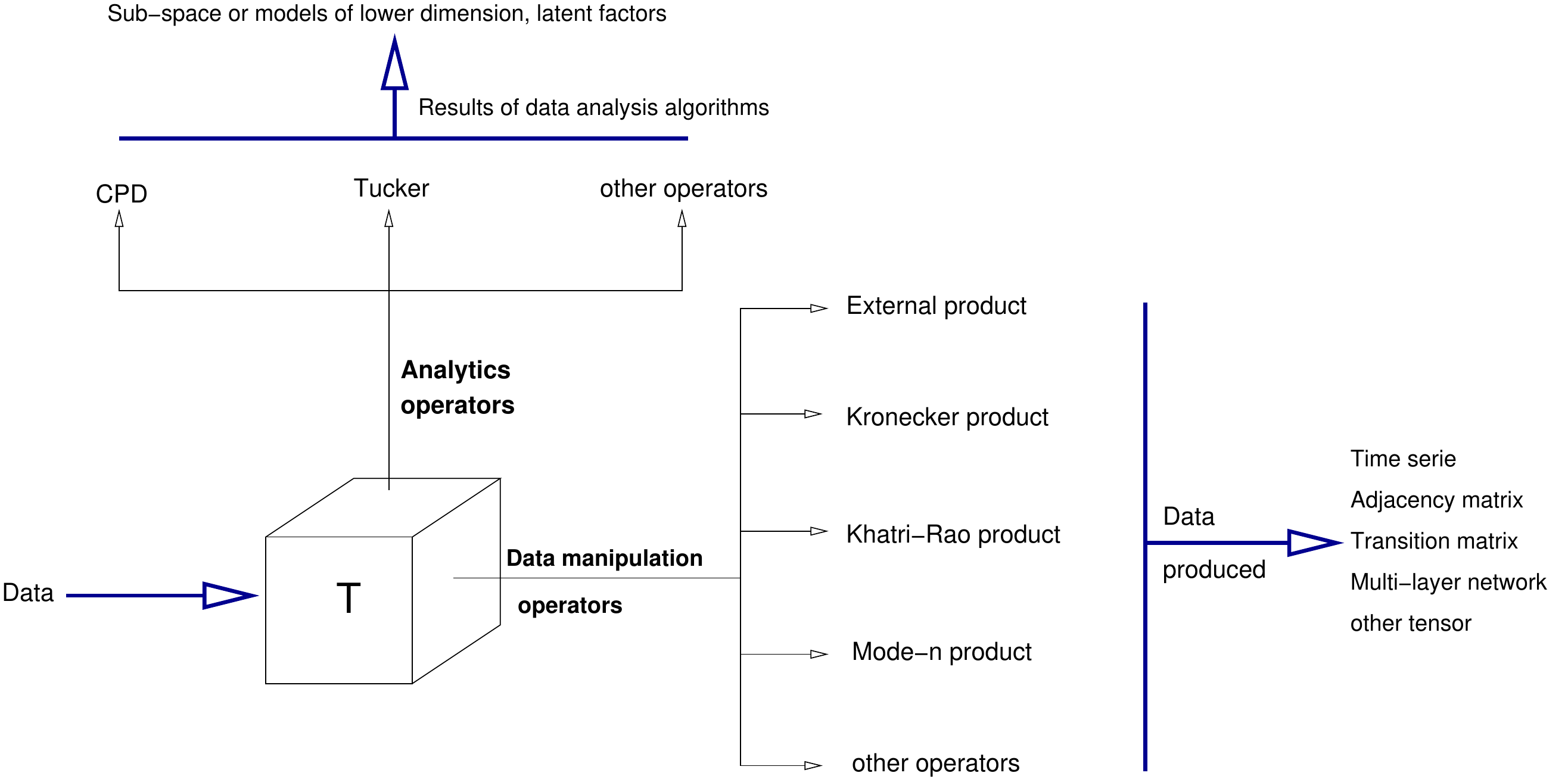}
\caption{Tensor Model and operations}
\label{fig:TDM}
\end{figure*}

Our objective is to provide the mathematical object tensor with operators to perform data manipulation (transformations) and analysis as well to set up the notion of schema and views to build a real tensor data model (figure \ref{fig:TDM}). Moreover, as tensor will play of role of pivot among different data models, we will define constructs in the model to support links to data sources.

\subsection{Architecture}

\begin{figure}[h!]
\centering
\includegraphics[width=7.5cm]{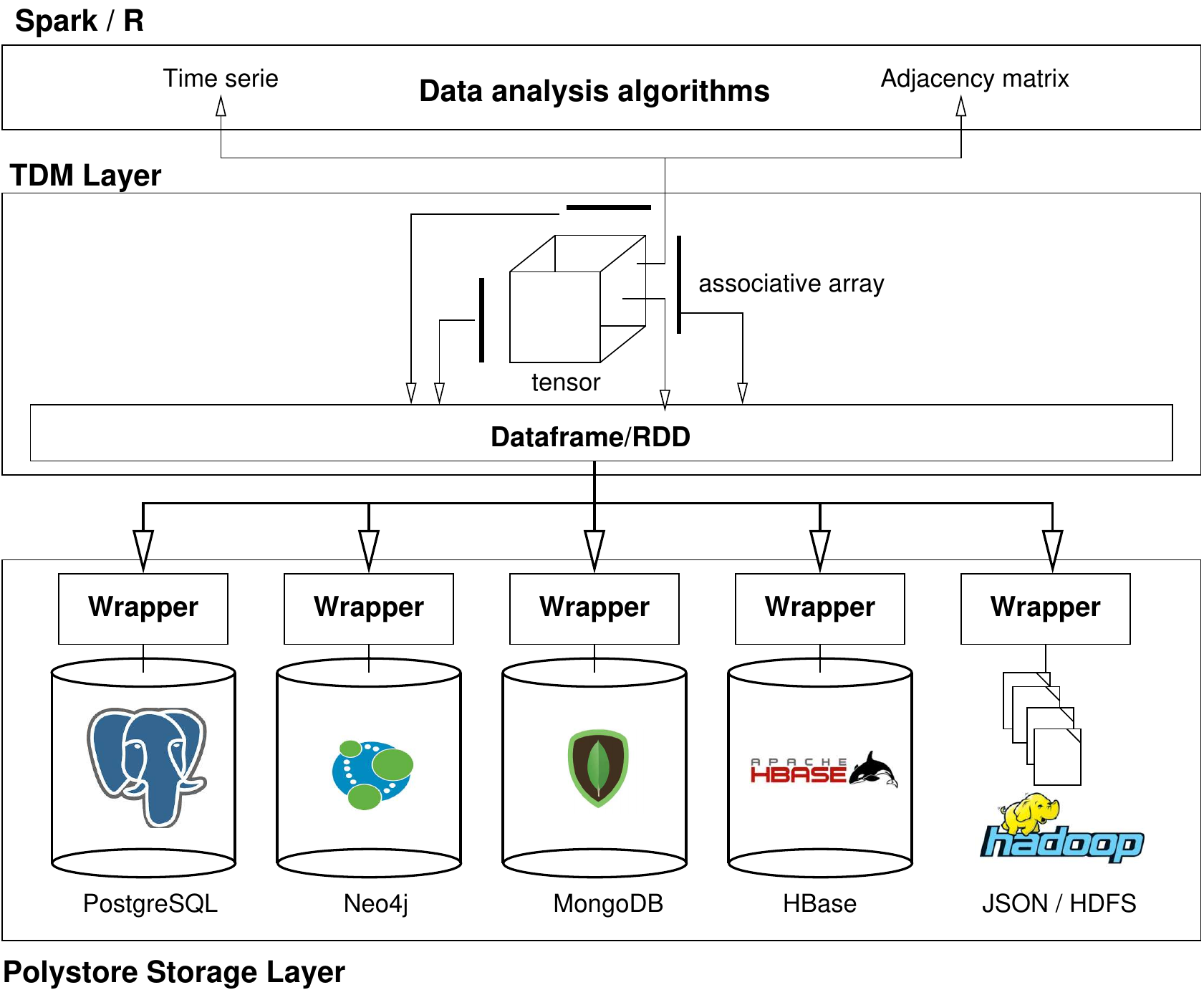}
\caption{Tensor and associative arrays roles in the architecture of a polystore system}
\label{fig:archi}
\end{figure}

The figure \ref{fig:archi} describes how data which fill tensors values are obtained from \textit{wrappers} which express the queries in the native language of each data store.

Queries for tensor construction are submitted to the \textit{wrappers} and have the same shape: they send back $N+1$ attributes where $N$ first attributes are the dimensions and the last one serves as value for the elements of the tensor (obtained with \texttt{GROUP BY}-like queries  on the attributes which represent the dimensions).  This feature allows us to implement \textit{wrappers} having all the same structure and so to simplify the models transformations. 
For R language, the \textit{wrappers} are implemented using the packages  R DBI, RNeo4j\footnote{\url{https://github.com/nicolewhite/RNeo4j}}, RMongo, RCassandra et RHBase\footnote{\url{https://github.com/RevolutionAnalytics/rhbase}}. In Spark, we work with the SQL layer, data frame and RDD (Resilient Data Sets) abstractions. To represent indexes used in each dimension of tensors we use associative arrays, their values are sets of identifiers (unique keys) that map sets of values of a specific data type to natural numbers, for example to associate each Twitter users, or hashtag to a natural number. Associative arrays are maps from $K \rightarrow \mathbb{N}$ where $K$, a dimension, is a set of atomic types (real, integer, string, etc.).
Associative arrays are translated into specific queries, sent to a storage system and materialized in the tensor model layer (figure \ref{fig:archi}). In Spark, they are then used in a multi-sources join to obtain values for a tensor. In R, associative arrays are stored as specific structure in the data source from which values are retrieved.

\section{\uppercase{TDM: A Tensor Data Model}}\label{sec:TDM2}

In this section, we briefly recall motivations and works around tensors in databases and complex networks fields. Then we present our tensor data model (TDM) and give some illustration examples.

\subsection{TDM Motivations}

In a database context, tensors are rather multidimensional arrays~\cite{Baumann2011}. For example, in SciDB \cite{Stonebraker2013}, the data model is based on arrays where each  cell can  be a  vector of values,  dimensions can  be  either  integers  or user-defined types. These models are useful for data produced in domains such as earth-sciences that manipulate matrices and time series.

Complex networks, and network science field of research aim at providing tools to model the structural complexity of a variety of complex systems such as transportation, biology, communication networks and online social networks. Although adjacency matrices are popular structures to model networks, such representation is insufficient for representing heterogeneous multi-layer networks. Real complex networks are heterogeneous i.e., nodes and edges have different types and edges types can be semantically grouped.
Achieving a deep understanding of such complex network data requires generalizing the traditional network theory. 
The concepts of multi-layer networks, multiplex networks, multi-relational networks or network of networks have been introduced recently to provide an expressive model for real-world complex networks \cite{Kivela2014}. 
In \cite{DeDomenico2013} and  \cite{Kivela2014} the authors use adjacency tensor as a model for multiplex networks\footnote{A multiplex network is a special type of multi-layer network in which the only possible types of inter-layer connections are ones in which a given node is connected to its counterpart nodes in the other layers.} and study popular analysis measures such as  degree of centrality, eigenvector centrality, clustering coefficients, random walks, modularity on multiplex network representation. They also discuss the relationships among multiple models such as graph, multigraph, hypergraph and linear and multi-linear algebras. In \cite{Rodriguez2010} the authors use a 3-order tensor as a model of  multi-relational network. One of the dimensions is used to represent the layers i.e. the different types of relationships, for example if there are $n$ nodes (individuals) and $r$ relationships between nodes (professional, family, friendship) the size of the 3-order tensor will be $n \times n \times r$ and $ijk$-i-th element has 1 for value if the node $i$ maintains a relationship of type $k$ with the node $j$, each relationship corresponding to a front slice of the tensor.
Multi-layer networks and their representation in tensors allow to model complex relationships among different dimensions, without having a fine knowledge and then understanding their links by means of an appropriate decomposition.

%
%
\subsection{TDM Formalization}

In TDM, tensor dimensions are represented by associative arrays. In the general case, an associative array is a map from a key space to a value space.  

\begin{definition}[Associative Array]
\label{def:AA}
An associative array is a map that associates keys to values as:
$$ A: K_1 \times \dots \times K_N \rightarrow \mathbb{V}$$
where $K_i, i=1,..,N$ are the sets of keys and $\mathbb{V}$ is the set of values. 
\end{definition}

The definition given in \cite{Kepner2017} restricts  $\mathbb{V}$ to have a semi-ring structure and the associative array to have a finite support. In TDM we use associative arrays for two cases.  First, we use different associative arrays denoted by $\mathbf{A}_i$ for $i=1,..,N$ to model dimensions of a tensor $\mathbfcal{X}$, in this case the associative array has only one set of keys associated with natural numbers $\mathbf{A}_i : K \rightarrow \mathbb{N}$. Second, an associative array is used to represent the values in each tensor. A $N$-order tensor $\mathbfcal{X}$ maps keys (n-uples) to a space of values (string, real, integer, etc.). 

\begin{definition}[Named Typed Associative Array]
\label{def:NTAA}
A named and typed associative array of  a tensor $\mathbfcal{X}$ is a triple $(Name, \mathbf{A}, T_A)$ where $Name$ is a unique string which represents the name of a dimension, $\mathbf{A}$ is the associative array, and $T_A$ the type of the associative array i.e. $K\rightarrow\mathbb{N}$. 
\end{definition}

According to the previous definition, the signature of a named typed associative array is $Name : K \rightarrow \mathbb{N}$.

\begin{definition}[Typed tensor]
\label{def:TT}
A typed tensor $\mathbfcal{X}$ is a tuple $(Name,D_A, V, T)$ where:
\begin{itemize}
\item $Name$ is the name of the tensor;
\item $D_A$  a list of named typed associative arrays i.e., one named typed associative array per dimension; 
\item $V$ is an associative array that store the values of the tensor;
\item $T$ is the type of the tensor, i.e. the type of its values.
\end{itemize}
\end{definition}


$V$ handles the sparsity of tensors.  Sparse tensors have a  default value (e.g. 0) for all the entries that not explicitly existing in the associative array. The signature of a typed tensor is $Name : D_A \rightarrow T$. A TDM schema is a set of typed tensors signatures.

\subsection{Examples with Twitter Data}

Even if Twitter data seem to have a simple structure they are actually very rich. Their corresponding relational schema  has only few relations for example, some tables are used to represent entities (i.e. tweets, hashtags, etc.), or social relationships (i.e. followers, etc.), foreign keys and association tables represent the use of operators such as RT, @, \#, URL. Nevertheless, the relational schema is not so easy to process because most analytic queries require multiple joins and auto-joins. Moreover, they usually contain complex group by clauses, sometimes with timestamps or ranking. Finally, some queries are recursive or require to compute the transitive closure of relations.

The social data retrieved from Twitter are by nature a multiplex network where the nodes are heterogeneous (users, tweets, hashtags, etc.) and the relationships too (retweet, publish, follow, mention, etc.). Moreover, different dimensions are existing such as users interactions (retweet, follow), users actions (publish, like), tweet structure (content, mention, hashtag, URL), so a richer model like multi-layered network model \cite{Kivela2014} should be used to perform meaningful analysis.

To illustrate the potential use of tensor, let start with a multi-layer network defined as $GM=(V,E,L)$ where $V=\{V_1, V_2, \dots, V_n\}$ is a partitioned set of nodes, $E=\{E_1,E_2,\dots, E_k\}$ is partitioned set of edges, with $E \subseteq V \times V$ and $E_i \subseteq V_l \times V_m$ for $i \in \{1,\dots, k\}$ and $l,m \in [1,n]$. $L$ is a partitioned set of layers, $L=\{L_1, L_2, \dots, L_p\}$ where $L_i\subseteq{E}$, with $L_i \cap L_j = \emptyset, \forall i,j$ modeling the dimensions.


Binary relationships are subsets of $E\times E$ and their associated functions $R_i$ can be used to associate values to edges or to count the number of edges between vertices. Binary relationships can be represented by matrices or tensor slices.  For example, $V_1$, $E_1$ and $R_1$ can model users and mentions (@ operator) and the number of mentions between two users. 
But all relationships in the social data do not have the same signature. Let us denote the set of users by $V_1$, the set of tweets  by  $V_2$ and take from example different types of relationships that do not have the same signature:
\begin{itemize}
 \item $mention$, $R_1: V_1 \times V_1 \rightarrow \mathbb{N}$ 
 \item $retweet$, $R_2: V_1 \times V_2 \rightarrow \mathbb{N}$ 
 \item $retweet_{U}$, $R_3: V_1 \times V_1 \rightarrow \mathbb{N}$, i.e. aggregated retweets according to source user 
 \item $publish$, $R_4: V_1 \times V_2 \rightarrow \mathbb{N}$
 \item $follow$, $R_5:  V_1 \times V_1 \rightarrow \mathbb{N}$
\end{itemize}

Let's look in more details the relationships $mention$, $retweet_U$ and $follow$, their associated tensor is  $\mathbfcal{T}$ from $V_1 \times V_1 \times \{R_1, R_3, R_5\} \rightarrow \mathbb{N}$. For example $\mathbfcal{T}(u_1,u_2,R_1)=5$ if the user $u_1$ has mentioned five times the user $u_2$ in its tweets.  Thus, a relationship between users can be modeled by a 2-order tensor and the set of users for their relationships  will be represented by a 3-order tensor\footnote{If all the relationships to be represented are homogeneous, that is to say if the associated functions have the same signature.}.

\begin{figure}[!h]
\centering
\begin{tikzpicture} [scale = 3]
\path[shape= coordinate]
(0,0,1)  coordinate(a)
(1,0,1) coordinate(b)
(1,0,0) coordinate(c)
(0,0,0) coordinate(d)
(0,0.75,1)  coordinate(e)
(1,0.75,1) coordinate(f)
(1,0.75,0) coordinate(g)
(0,0.75,0 ) coordinate(h);
\draw[style=dashed] (d)--(c);
\draw (c)--(g)--(h);  \draw[style =dashed] (h)--(d);   
\draw(c)--(b)--(a) node[below, pos=0.5, sloped] {\textit{tweet (j)}};  
\draw[style =dashed] (a)--(d);
\draw(g)--(f)--(e)--(h) node[above, pos=0.5, sloped] {\textit{time (k)}};
\draw(e)--(a) node[below, pos=0.5, sloped] {\textit{user (i)}};    \draw(f)--(b);
\draw(1,0.1,1)  node[left] {1};
\draw(0.1,0.75,1) node[below] {1};
\draw(0.75,0.75,1) node[below] {1};
\draw(0.5,0.5,1) node[below] {1};
\draw(0.92,0,-0.2) node[left]{\color{lightgray}{1}};
\draw(0,0.75,0) node[below right]{\color{lightgray}{1}};
\draw(0.4,0.6,0) node[below] {\color{lightgray}{1}};
\draw(0.75,0.3,0) node[below]{\color{lightgray}{1}};
\end {tikzpicture}
\caption{Associated tensor $\mathbfcal{X}$}
\label{fig:tpublish}
\end{figure}
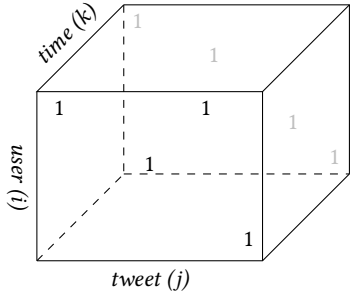

Let's now take another example with three sets of nodes $users$ defined by $V_1$, $tweets$ defined by $V_2$ and $time$ defined by $V_3$  for  representing the relationship $publish$, i.e. a user publishes a tweet on a given day. The different named typed associative arrays are given on figure \ref{fig:aa}. Figure \ref{fig:tpublish} depicts the tensor $\mathbfcal{X}$ and its values. The value 1 at coordinate (1,3) in the front slice means that the user u3 has posted the tweet t1 on day 18-03-08.



 However, it is not so easy to model the presence of several operators in the same tweet, for example the co-occurrence of hashtags can be used to explain the meaning of the first one if it is very general term and a mention can be added at the beginning of the tweet as a call to an answer. These kinds of specific relationships can be modeled by using hyperedge in an hypergraph that can also be transformed into a tensor that encompasses both simple relationship and complex ones like $\mathbfcal{T}: V_1 \times V_2 \times V_4 \times V_4 \times V_1 \rightarrow \mathbb{N}$, where $V_4$ is the hashtags used in the tweets.
 

This example highlights the following elements:
\begin{itemize}
\item When a user writes only one mention in a tweet without hashtags, the dimensions that represent hashtags should also include a $null$ value. This $null$ value is easily supported by introducing a specific value in associative arrays;
\item As the order of tensor increases the sparsity also increases, at first it seems to be natural to define or adapt the theory of normal forms from relational data model to tensor\footnote{In the previous example it will probably drive us to define 3 or more 2-order tensor for users and hashtags, users and mentions, users and tweets.}. Using the $null$ value a tensor can model both simple relationships and more complex relationships this can also contribute to gain in performance by allowing materialized joins.  
\end{itemize}

\begin{figure}[!h]
\begin{center}
\begin{tabular}{c|c|c|c|c}
  user & u1 &  u2 & u3 & $\dots$ \\ \hline
i  & 1 & 2 & 3 & $\dots$
\end{tabular}

\medskip

\begin{tabular}{c|c|c|c|c|c}
  tweetID & t1 &  t2 & t3 & t4 &$\dots$ \\ \hline
 j & 1 & 2 & 3 & 4 & $\dots$
\end{tabular}

\medskip

\begin{tabular}{c|c|c|c|c|c}
  time & 18-03-08 & 18-03-07 & 18-02-28 & 18-02-26 &$\dots$ \\ \hline
 k & 1 & 2 & 3 & 4 & $\dots$
\end{tabular}
\end{center}
\caption{Named Typed Associative Arrays representing tensor dimensions}
\label{fig:aa}
\end{figure}

\section{\uppercase{TDM's Operators} \label{sec:op}}

To carry out a wide range of queries it should possible to define several of the standard operators from relational algebra in terms of tensor operations. In \cite{Hutchison2017,Kepner2017} the authors define a model and operators over associative arrays  to unify relational, arrays, and key-value algebras. Our operators are defined to provide programmers with a logical data independence layer i.e. to bridge the semantic gap between analysis tools and storage systems. Our set of operators works at two different levels: at the associative array level and at the tensor level.

\subsection{Data Manipulation Operators}

A fiber of a tensor $\mathbfcal{X}$ is  a vector obtained by fixing all but one $\mathbfcal{X}$'s
indices: $\mathbfcal{X}_{:jk}$, $\mathbfcal{X}_{i:k}$ et $\mathbfcal{X}_{ij:}$. Fibers are always assumed to be column vectors, analogue of matrix rows and column.
 A slice of a tensor $\mathbfcal{X}$ is a matrix obtained by fixing all but two of $\mathbfcal{X}$'s indices: $\mathbfcal{X}_{i::}$, $\mathbfcal{X}_{:j:}$ et $\mathbfcal{X}_{::k}$.

A project operator can be generalized by the Hadamard product of a $N$-order tensor with a boolean tensor of the same order that contains $1$  for the elements to be selected: $\mathbfcal{X}\circledast\mathbfcal{B}$. 

For example, for a 3-order tensor, $\mathbfcal{X}_1$, representing users, hashtags used (their number of occurrences in the tweets of a user) and the time, to select all hashtags used by a user $u_i$, the result will be in a 2th-order tensor such as: $\mathbfcal{X}_2=\mathbfcal{X}_1\circledast\mathbfcal{B}_1$ with $\mathbfcal{B}_{1_{i::}}=1$. To obtain a time series reflecting the use of a hashtag, the sum of the columns of the 2th-order tensor obtained is carried out.


\bigskip

A select operator can act on two levels: 1) on the values contained in the tensor (equivalent to a selection on a single relational attribute) or 2) on the values that are in associative arrays  $\mathbf{A}_i$, $i=1,..,N$.  The select operator $\sigma$ is written with two conditions, the first on the dimensions, the other on the values: 
$$\sigma[cdt\ dim][cdt\ val]\mathbfcal{X}$$
The condition on the dimensions is made by the product of Hadamard with Boolean tensor whose elements which have 1 for  value correspond to the elements of the $\mathbf{A}_i$ selected.
The following example selects tweets published by the user $u1$ and time is comprising between 18-03-08 and 18-02-28, from the tensor $\mathbfcal{X}$:
$$\sigma[U='u1'\wedge T\geq '18-02-28' \wedge T\leq '18-03-08'][=1]\mathbfcal{X}$$

\subsection{Analytical Decomposition Operators}
\begin{figure}
\centering
\includegraphics[width=7.5cm]{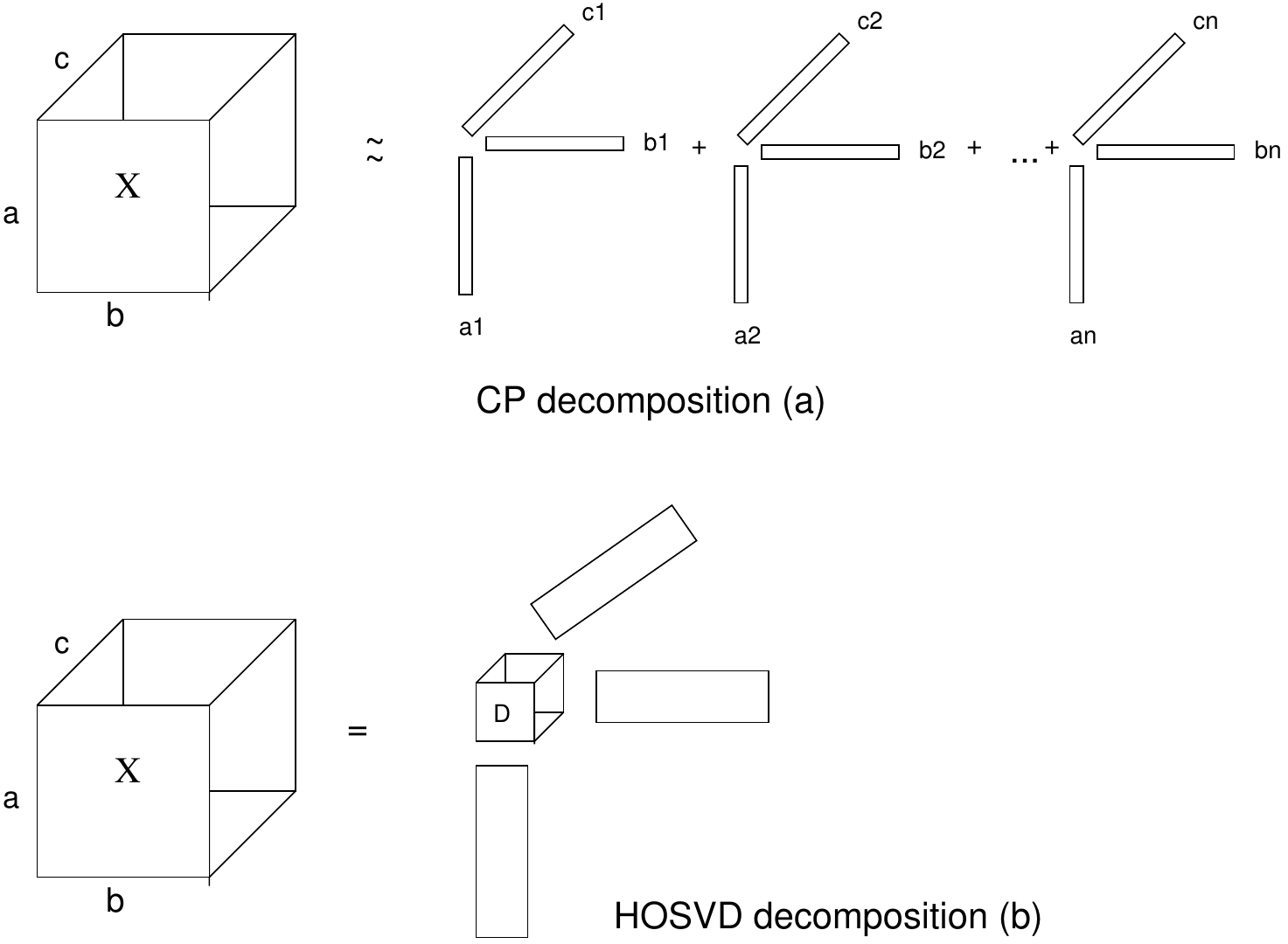}
\caption{CP and HOSVD decompositions}
\label{fig:decom}
\end{figure}

Tensor decompositions such as CANDECOMP/PARAFAC (CP), Tucker, HOSVD are used to perform dimensionality reductions and to extract latent relations \cite{Kolda2009}. 
Since tensor representations of  data are multiple and their semantics are not explicit, the results of tensor decompositions are complex to interpret. However, by analogy with the matrix decompositions it is possible to determine the decomposition associated with an objective, or to answer to a question, by using a proper tensor model. For example, to express each space generated by one of the dimensions of the tensor independently of the others, but depending on the global space, we can use a decomposition CP (figure \ref{fig:decom}(a)). To determine user models based on hashtag or recurring patterns of behavior we prefer to use a HOSVD decomposition (figure \ref{fig:decom}(b)). Then models produced by HOSVD decomposition can be used in recommendation systems.

\section{Scientific Application Use Case}
\label{sec:exp}
In this section, we validate our approach by a proof-of-concept, showing how TDM is used in the context of a real project (TEP 2017) involving collaborations with social scientists. We will not dwell here on the results of the interpretations made by social scientists. 

The data we are working on, are a part of 50M tweets corpus collected during the french presidential election in 2017. Data are stored in JSON file format in HDFS, most important attributes of tweets are stored in a relational database in a unique table and in another database with a schema in third normal form. The main research objective of this project is to better understand where and how discourse relating to emerging political issues circulates on the French Twitter-sphere, during presidential election campaign in 2017.  

We choose to focus our study on virality, the initial list of potential viral tweets (i.e. tweets that have been propagated a lot) reduces the corpus of 50M of tweets to thousand, giving the possibility to social scientists to validate our experiments by a qualitative study.
According to Bessi and al. \cite{bessi2016}, robots played an important role during the American presidential election of 2016. The authors estimated that about 400,000 robots are engaged, responsible for roughly 3.8 million tweets (19\% the entire conversation). 

Our objective is to understand how bots artificially relieved tweets and to analyze this phenomenon. 
From the model point of view and more precisely, the construction of the tensor, we wish to observe the quantity and complexity of code necessary to transform the data in two cases: 1) using R directly for the analyses; 2) using the TDM model like intermediate. 

\subsection{Viral Tweets}
Virality can be defined by three parameters summarized by  Beauvisage and al. \cite{beauvisage2011}:  temporal concentration of the attention to the content,  traffic of this content and  mechanisms of the contagion from an individual to the other one. Unlike tweets that make buzz, the start of their broadcast is slower and it lasts longer in time.
Various metrics is taken into account to determine tweet virality  \cite{hoang2011,ma2013,weng2014}. We have lexical metrics with study of the contents  (presence of URL and hashtags, construction of the hashtag from several words, etc.) and  contextual metrics with the activity of the account, its community, etc. as well as the time. Number of  metrics makes difficult the interpretation of obtained results.

We reduced the global corpus to the period  between two ballots of the election (from April 24th, 2017 till May 7th, 2017) and calculated the number of retweets for every tweet. A sample of the most popular tweets that is, in our case, the most propagated by retweets was obtained by selecting tweets retweeted at least 1 000 times over the period. This sample contains 1,123 tweets among which some were retweeted more of 20 000 times. The list of tweets id is available on GitHub\footnote{\url{https://github.com/EricLeclercq/TEE-2017-Virality}} for a reproducibility of results. 
For each tweet, we then studied the time series for the frequency i.e. the number of retweets by period of one hour, 4 hours, 8 hours, 12 hours and 24 hours, calculated the speed of propagation and extracted, by use of the algorithm breakout\footnote{The algorithm is based on the method E-Divisive with Medians (EDM), it uses a statistical measure of energy to detect differences of the average \url{https://github.com/twitter/BreakoutDetection}.}, the intervals of time in  which activity was important. 

Queries which produced the analysis results are expressed in the native language of the storage system, here SQL. In both cases, queries are launched from R on the PostgreSQL database with normalized schema and produce data for the algorithms. It takes in less than a minute to produce tensor dimensions and less than 5 minutes to produce tensor values. It is approximately the same execution time for build the dataset without using the tensor. In the tensor (users, hashtags, time) case the code is divided in several small queries (dimensions, values). In the other case a unique, complex to read, query with more than 15 lines of SQL produces the data set.


\subsection{Bots Detection}

In order to understand the diffusion mechanisms  of supposed viral tweets, we first studied the share of activity related to bots or rather accounts whose behavior is similar to  humans.

Several methods have been proposed to detect robots \cite{varol2017}, they most often aggregate a large number of features to produce a predictive model based on a learning algorithm such as \textit{random-forest}. An experiment using the OSoMe API\footnote{\url{https://botometer.iuni.iu.edu/}} to obtain a probability of automated behavior (of robot type) leds us to note that the values of the probabilities were not enough significant  to detect bots during the studied period.  One of the assumptions is that it is hybrid accounts of users assisted by algorithms. However, simple criteria such as the maximum number of tweets published in one hour makes it possible to unambiguously find some accounts with automated behavior, confirmed by the manual study, which will serve as a marker for the other analyzes.
Based on the observation that real (human) accounts publish tweets on a regular basis, basic statistics such as the average of tweets or retweets sent per hour do not make it easy to extract robots.

Bots do not retweet randomly, so from the tweets contents (hashtags)  we built a 3-order tensor modelling the users $U$ having retweeted a supposed viral tweet, the hashtags $H$ contained in these tweets and the time $T$ (14 days of observation). We got a tensor containing potentially $1,077 \times 568 \times 336$ items; the values of the tensor therefore represent the number of occurrences of each hashtag per user per hour by considering only the retweets of supposed viral tweets. 
The research space being very large, then we performed a CP decomposition to reduce the user space based on their behavior. The decomposition CP produces $n$ groups of three 1-order tensors, here vectors $U$, $H$, $T$ (see figure \ref{fig:decom}(a)). We then apply the k-means clustering algorithm to identify groups of users. The retained value of $n$ is the one from which there is no more modification of the clusters. Experimentally, we get $n=8$  and therefore a user is described by a point in an 8-dimensional space. The k-means algorithm applied to this data  determines 4 groups of users:  a group of one account already detected as a robot, a group of two accounts, a group of about thirty accounts comprising more than half users with a probability of being a robot greater than 0.6 and a last group containing other users. The group of two accounts, revealed after manual study, to be linked (same behavior and hashtags) and assisted by an algorithm that retweets messages against the Macron candidate. These accounts had evaded other analysis techniques.

The tensor construction from the data and the tensor decomposition in R take each less than 5 minutes, a thorough study of performance will be required.

\section{Conclusion}
\label{sec:conclusion}
In this article,  we have proposed a new architecture for social networks data warehousing based on a polystore system and a tensor-based pivot data model. The tensor model makes it possible to generalize matrix representations (adjacency matrix, time series, etc.) as well as graphs including multigraphs and hypergraphs and  to take into account models of complex networks (multi-layer networks, multi-relational networks, etc.). We also presented some data manipulation and analysis operators on the tensor model.

The work has been experimented with Twitter data. We  detected viral tweets by focusing on time series. In order to understand the information dissemination mechanism, we also studied the significance of social bot activity in diffusion. Our results have been validated by social scientists and researchers in communication sciences.  This experiment demonstrated the tensor modeling capabilities and the relevance of the architecture according to the ease of implementation of model transformation in analyses.

The perspectives concern the complete formalization of a set of operators as well as the study of the properties of the algebraic structure that they generate (semi-ring for example). In parallel, we want to develop a real prototype of architecture in order to study queries optimization including tensor operators. Nevertheless, semi-ring structures give operators good properties for distributed implementation, which suggests a good potential for scaling up.

\bibliographystyle{abbrv}
\bibliography{./Biblio/biblio.bib}
%

\end{document}